\pdfoutput=1
\RequirePackage{fixltx2e}
\documentclass[aip,jcp,reprint,floatfix]{revtex4-1}

\usepackage{amsmath,amsfonts} %
\usepackage{wasysym} %
\usepackage[acronym]{glossaries} %
\usepackage{graphicx} %
\usepackage{dcolumn} %
\usepackage{bm} %
\usepackage{ifpdf} %
\usepackage{color} %
\usepackage{nicefrac} %
\usepackage{natbib} %
\usepackage[byname]{smartref} %
\usepackage[breaklinks, %
colorlinks, %
linkcolor=blue, %
citecolor=blue, %
urlcolor=blue]{hyperref} %
\usepackage[table]{xcolor} %
\usepackage{ulem} %

\DeclareMathOperator{\tr}{Tr} %
\DeclareMathOperator\diag{diag} 
\newcommand{\eq}[1]{Eq.~(\ref{#1})} %
\newcommand{\bea}{\begin{eqnarray}}
\newcommand{\eea}{\end{eqnarray}}
\newcommand{\ket}[1]{\ensuremath{\left|#1\right\rangle}}
\newcommand{\braket}[2]{\ensuremath{\left\langle #1|#2\right\rangle}}
\newcommand{\braOket}[3]{\ensuremath{\left\langle #1\left|#2\right|#3\right\rangle}}

\newcommand{\mat}[1]{\ensuremath{\boldsymbol{#1}}}
\newcommand{\inumb}[0]{\dot{\imath}}
\newcommand{\Dim}[0]{\mathcal{D}}

\ifpdf %
 \fi
                            
\newacronym{CI}{CI}{conical intersection} %
\newacronym{DOF}{DOF}{degrees of freedom} %
\newacronym{PES}{PES}{potential energy surface} %
\newacronym{RMSD}{RMSD}{root mean square deviation} %
\newacronym{FC}{FC}{Franck-Condon} %
\newacronym{FAD}{FAD}{flavine adenine dinucleotide} %
\newacronym{HDF}{HDF}{7,8-didemethyl-8-hydroxy-5-deazariboflavin} %
\newacronym{PL}{PL}{photolyase} %
\newacronym{CPD}{CPD}{cyclobutane pyrimidine dimer photoproduct} %
\newacronym{ET}{ET}{electron transfer} %
\newacronym{64PP}{64PP}{pyrimidine-pyrimidone (6-4) photoproduct} %
\newacronym{NFGR}{NFGR}{non-equilibrium Fermi golden rule} %
\newacronym{CASSCF}{CASSCF}{complete active space self consistent field} %
\newacronym{CASCI}{CASCI}{complete active space configuration interaction} %
\newacronym{XMCQDPT2}{XMCQDPT2}{extended multiconfiguration quasi degenerate second order perturbation theory} %
\newacronym{HOMO}{HOMO}{highest occupied molecular orbital} %
\newacronym{LUMO}{LUMO}{lowest unoccupied molecular orbital} %
\newacronym{QVC}{QVC}{quadratic vibronic coupling} %
\newacronym{LBM}{FB}{flavin bending} %

\newcommand{\ip}{{pri}} 
\newacronym{pm}{primary mode}{primary mode} %
\newacronym{ps}{primary system}{primary system} %
\newacronym{5D}{5D}{five-dimensional} %

\newcommand{\is}{{sec}} 
\newacronym{sm}{secondary mode}{secondary mode} %
\newacronym{ss}{secondary system}{secondary system} %

\newcommand{\ifull}{{ful}} 

\begin{document}

\title{Nuclear dynamics investigation of the initial electron transfer in the cyclobutane pyrimidine 
dimer lesion repair process by photolyases} %

\author{Lo{\"i}c Joubert-Doriol} %
\affiliation{Department of Physical and Environmental Sciences, University of Toronto Scarborough, Toronto, Ontario M1C 1A4, Canada} %
\affiliation{Chemical Physics Theory Group, Department of Chemistry, University of Toronto, Toronto, Ontario M5S 3H6, Canada} %
\affiliation{Dipartimento di Chimica, Università di Siena, via De Gasperi 2, I-53100 Siena, Italy} %
\author{Tatiana Domratcheva} %
\affiliation{Department of Biomolecular Mechanisms, Max Planck Institute for Medical Research, Jahnstrasse 29, 69120 Heidelberg, Germany}
\author{Massimo Olivucci} %
\affiliation{Dipartimento di Chimica, Università di Siena, via De Gasperi 2, I-53100 Siena, Italy} %
\affiliation{Department of Chemistry, Bowling Green State University, Bowling Green, OH 43403, USA} %
\author{Artur F. Izmaylov} %
\affiliation{Department of Physical and Environmental Sciences, University of Toronto Scarborough, Toronto, Ontario M1C 1A4, Canada} %
\affiliation{Chemical Physics Theory Group, Department of Chemistry, University of Toronto, Toronto, Ontario M5S 3H6, Canada} %

\date{\today}

\begin{abstract}
Photolyases are proteins capable of harvesting the sunlight to repair DNA damages caused by UV light. In this work we focus on the first step in the repair process of the \gls{CPD} lesion, which is an \gls{ET} from a flavine cofactor to \gls{CPD}, and study the role of various nuclear \gls{DOF} in this step.
The \gls{ET} step has been experimentally studied using transient spectroscopy and the corresponding data provide excellent basis for testing the quality of quantum dynamical models.
Based on previous theoretical studies of electronic structure and conformations of the protein active site, we present a procedure to build a 
diabatic Hamiltonian for simulating the \gls{ET} reaction in a molecular complex mimicking the enzyme's active site.
We generate a reduced nuclear dimensional model that provides a first non-empirical quantum dynamical 
description of the structural features influencing the ET rate. 
By varying the nuclear \gls{DOF} parametrization in the model to assess the role of different nuclear motions,
we demonstrate that the low frequency flavin butterfly bending mode slows \gls{ET} by reducing 
Franck-Condon overlaps between donor and acceptor states and also induces decoherence.
\end{abstract}

\maketitle

\glsresetall

\section{Introduction}
\label{sec:introduction}


UV radiation causes damage to DNA by forming bonds between contiguous thymine bases. 
There are two types of such lesions: the \gls{CPD} and the \acrlong{64PP}. 
Both lesions can be repaired by flavoproteins: \glspl{PL} that are capable of splitting the thymine dimers back to monomers.~\cite{Sancar:2003/cr/2203,Sancar:2008/jbc/32153,Muller:2009/cosb/277}
While \glspl{PL} are not present in Humans, they have been found in bacteria, fungi, plants and animals,~\cite{Sancar:2003/cr/2203} these proteins proved to be useful in protection against sunburns and skin cancer prevention when added to sunscreens.~\cite{Berardesca:2012/mmr/570,Emanuele:2013/jdd/1017,Emanuele:2014/jdd/309} 
Understanding the molecular-level details of the repair processes is of high interest not only for our basic knowledge, but also for the potential improvement of the efficiency of such treatments or for drug design.
Theoretical models are required to unravel microscopic mechanisms at work in the \gls{PL}s' active site by comparison of the model predictions with the available experimental measurements. In this paper we present a methodology to construct such quantum dynamical models.


The action of \gls{CPD}-\gls{PL} is understood as follows:~\cite{Mees:2004/science/1789,Kao:2005/pnasusa/16128}
After harvesting the sunlight~\cite{Weber:2005/bba/1,Essen:2006/cmls/1266,Zheng:2008/jpcb/8724}, 
the protein utilizes the energy excess to break the undesirable bonds between the thymine bases.
This process happens due to a cofactor present in the active site of \gls{CPD}-\gls{PL}: a \gls{FAD}. \gls{FAD} consists of riboflavin connected to adenosine by two phosphates. 

\begin{figure}
  \centering
  \frame{\includegraphics[clip,width=0.47\textwidth]{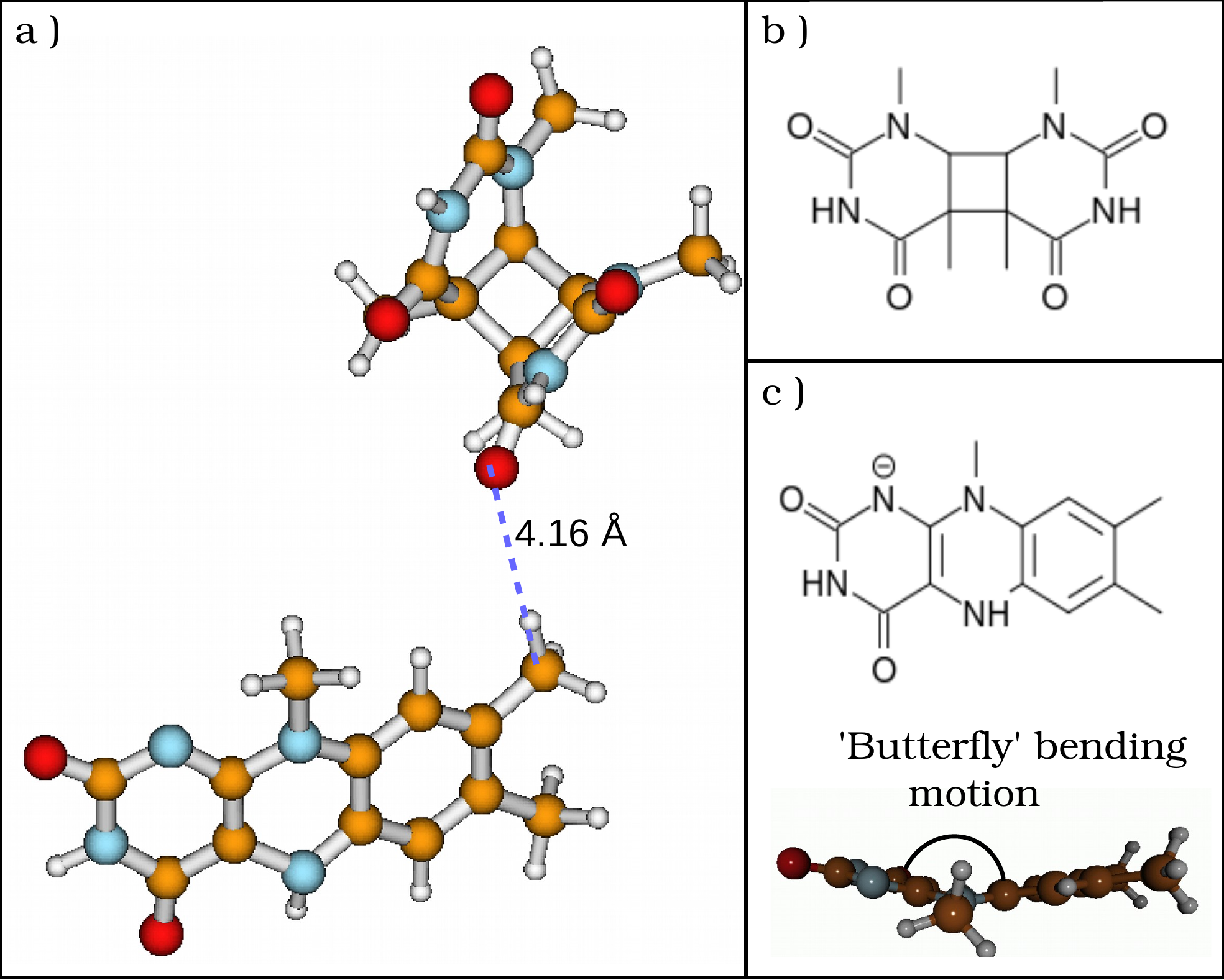}}
  \caption{a) Minimal molecular model for the \protect\gls{CPD}-\gls{PL} active site. b) Chemical structure of the \protect\gls{CPD}. c) chemical structure of lumiflavin in two-electron reduced state and definition of the 'butterfly' bending motion.}
  \label{fig:defs}
\end{figure}

Due to reduction of the isoalloxazine ring of riboflavin (lumiflavin in Fig.~\ref{fig:defs}), the \gls{FAD} cofactor is present in either two 
redox states: FADH$^-$ or FADH$^\bullet$.
The importance of the \gls{FAD} cofactor is in its role of electron donor to the pyrimidine dimer to initiate the repair process.
The repair process includes the following steps:~\cite{Kao:2005/pnasusa/16128} 1) light excitation of a flavin-type antenna cofactor, 
\acrlong{HDF}, 2) energy transfer to the \gls{FAD}H$^-$, 3) forward \gls{ET} from excited \gls{FAD}H$^-$ to the thymine dimer, 4) bond dissociations, 5) electron back transfer to \gls{FAD}H$^\bullet$.


The forward \gls{ET} is a population transfer between a donor state, where the electron is located on excited flavin, 
and an acceptor state, where the electron is located on \gls{CPD}.
The lifetime of the \gls{ET} reaction has been measured to be $170$ ps by transient absorption experiments at room temperature.~\cite{Kao:2005/pnasusa/16128,Liu:2012/jacs/8104}
Such a long timescale indicates a small coupling between the two electronic states, which is supported by the first estimate of this coupling in Ref.~\citenum{Antony:2000/jacs/1057}: $7.10^{-4}$ eV. In this case, the electron is effectively diabatically trapped in the donor state.\cite{Blancafort:2001/JACS/722,Blancafort:2005/jacs/3391} It was found recently 
that capturing quantum interference in the nuclear wavefunction can be crucial for adequate modelling of 
such systems.\cite{Joubert:2013/jcp/234103} 
Therefore, a fully quantum treatment of the nuclear motion using vibronic models appears to be necessary. 
However, building a vibronic model for systems of our interest leads to another major difficultie, which is the large 
number of nuclear \gls{DOF}, over one hundred that are involved in the \gls{ET} process. 
To address these challenges in an adequate and computationally feasible way
we use a perturbative approach, which is an extension to the usual Fermi Golden Rule:~\cite{Borrelli:2011/pccp/4420,Borrelli:2015/jctc/415} the \gls{NFGR}.~\cite{Izmaylov:2011/jcp/234106,Endicott:2014/jcp/034104}

\glsunset{pm}
\glsunset{sm}
\glsunset{ps}
\glsunset{ss}


The \gls{NFGR} method scales only cubically~\cite{Endicott:2014/jcp/034104} with the number of \gls{DOF} in the calculation of the time evolution of the donor state population and.
Therefore, \gls{NFGR} seems well suited to treat \glspl{ET} in large systems. There are two conditions for \gls{NFGR} success: 1) a small coupling between donor and acceptor states
for applicability of the perturbation theory, and 2) small amplitude motions, since the method employs the harmonic approximation with rectilinear coordinates for the nuclear coordinate. 
Here, both conditions are assumed to be satisfied. The first, because of the very long time scale of the \gls{ET}.
The second, because the protein environment constrains the relative 
motion of \gls{FAD}H$^-$ and \gls{CPD}.~\cite{Liu:2013/pnas/12972}
The \gls{NFGR} method requires a diabatic representation of the Hamiltonian~\cite{Izmaylov:2011/jcp/234106,Endicott:2014/jcp/034104} built based on the first-principles electronic structure calculations.
Here, we apply a diabatization by ansatz to minimize non-adiabatic couplings.~\cite{Cederbaum:1981/ijqc/251,Koppel:1984/acp/59,Joubert:2014/jcp/044301} 
We construct a diabatic vibronic Hamiltonian model describing the initial \gls{ET} step occurring 
in the \gls{CPD}-\gls{PL} complex. The model considers two electronic states: donor and acceptor. We employ 
multi-configurational quantum chemical methods to map the relevant \glspl{PES} of the \gls{CPD}-\gls{PL} complex~\cite{Domratcheva:2011/jacs/18172}, then we optimize 
model Hamiltonian parameters using the least-square method on electronic state energies.


By using the model outlined above, here we investigate the role of nuclear \gls{DOF} in the \gls{ET} process and study their impact on the \gls{ET} lifetime. 
We put a special emphasis on a \gls{LBM} motion that is commonly identified as the flavin butterfly motion 
and is defined in Fig.~\ref{fig:defs}-c.
Two-electron reduced lumiflavin is anti-aromatic and is bent in the middle dihydrodiazine ring where 
nitrogens have a $sp^3$ hybridization in the ground 
state,~\cite{Walsh:2003/theochem/185} while it is found to be planar in excited states. 
Therefore, the low-frequency \gls{LBM} is believed to play an important role in the non-adiabatic 
dynamics of the FADH$^-$ cofactor.~\cite{Kao:2008/jacs/13132} 
On the other hand, the protein environment may constraint the active site and prevent large amplitude 
motions such as \gls{LBM}.~\cite{Liu:2013/pnas/12972}
As detailed below, we observed that the \gls{LBM} motion allows for decoherence and slows down the \gls{ET} reaction by decreasing Franck-Condon overlaps between the donor and acceptor electronic states.


The rest of the paper is organized as follows. Section \ref{sec:methods} details the methodology.
Section \ref{sec:results} discusses main features of \glspl{PES}, model variations and \gls{ET} rate estimates. 
Section \ref{sec:conclusion} concludes the paper and gives an outlook for future applications.

\section{Methods}
\label{sec:methods}

\subsection{Molecular model and its electronic structure}
\label{sec:MolMod-elec}

A molecular model of the active site is taken from Ref.~\citenum{Domratcheva:2011/jacs/18172} 
and contains the FADH$^-$ electron donor, modelled by lumiflavin, 
and the electron acceptor, the \gls{CPD} lesion of two methylthymines (see Fig.~\ref{fig:defs}). 
The initial structure of the two moieties is taken from the \gls{CPD}-\gls{PL} repair compex 
crystal structure (PDB file 1TEZ~\cite{Mees:2004/science/1789}).
In order to simulate geometrical constraints of the protein environment, the nuclear coordinates of the model's terminal atoms 
 in the model are frozen during geometry optimizations. Thus, these coordinates are not 
considered in the \glspl{PES} calculations.
After taking into account the geometrical constraints, the final model contains $\Dim=153$ nuclear (vibrational) \gls{DOF}.
The coordinates of all frozen atoms are given in the supplemental materials.~\cite{SuppInfo}


Based on previous work in Ref.~\citenum{Domratcheva:2011/jacs/18172}, we use the principal orbital \gls{CASSCF} approach~\cite{Udvarhelyi:2011/pp/554} further improved by the \gls{XMCQDPT2}.~\cite{Granovsky:2011/jcp/214113}
In this approach, the selected active space contains only the 
orbitals and electrons necessary to describe the static electron correlation, 
while \gls{XMCQDPT2} accounts for the dynamical electron correlation.
At the ground state equilibrium geometry, the active orbitals are: the \gls{HOMO}, $\pi_f$, located on lumiflavin, the \gls{LUMO}, $\pi_f^*$, located on lumiflavin, and the antibonding $\pi_{CO}^*$ orbital of the closest carbonyl group located on the thymine dimer. This active space gives six electronic configurations.
The choice of the active space is guided by an assumption that involvement of out-of-plane deformations 
of the lumiflavin's $\pi$-system in the ET process is very limited. 
Thus, the three active orbitals will not change their character along the entire reaction path and, hence are sufficient for a balanced description of the \gls{ET} process. 
The remaining orbitals will mainly only contribute to the dynamic electron correlation energy. The basis set used for the calculations is 6-31G$^*$. All active space molecular orbitals are given in the supplemental materials.~\cite{SuppInfo}
Geometry optimization, gradient, and Hessian calculations are performed at the \gls{CASSCF} level with three-root state-average orbitals.~\cite{Granovsky:2015/jcp/231101} The resulting \gls{CASSCF} energies are then corrected via \gls{XMCQDPT2} computations.
Analytic gradients at the \gls{XMCQDPT2} level of theory are not yet implemented and numerical gradient would be too cumbersome for such a high dimensional system. 
For electronic structure calculations we use the Firefly quantum chemistry package,~\cite{firefly} which is partially based on the GAMESS~(US)~\cite{gamess} source code.

We refer to the ground electronic state $S_0$ as the closed shell anionic configuration where two 
electrons are localized 
on the lumiflavin $(\pi_f^2)$ orbital. The donor state is the FADH$^-$ excited state, labeled as $S_D$, 
where excess of negative charge is still localized on lumiflavin with the $(\pi_f\pi_f^*)$ 
dominant configuration. The acceptor state, labeled as $S_A$, where an extra electron is localized on 
\gls{CPD} (the electron transfer state), has the $(\pi_f\pi_{CO}^*)$ dominant configuration.

The states $S_D$ and $S_A$ are our quasi-diabatic states, which are coupled via potential terms, 
their crossing gives rise to a \gls{CI} between $S_1$ and $S_2$ in the
adiabatic representation.
The energy minimum of the \gls{CI} seam was found by energy minimization  
constraining the energy difference between states to be zero 
with the Lagrange multiplier method.~\cite{firefly}
Currently, there is no implementation of analytic non-adiabatic coupling elements 
for systems of our size, therefore, an approximate numerical approach was used 
to obtain these elements. The CASSCF 
wavefunctions, $\Psi_j$, were numerically differentiated considering them as if they were 
\gls{CASCI} wavefunctions, i.e. assuming frozen molecular orbitals
\bea\label{eq:numNAC}
\left\langle\Psi_1 \bigg| \frac{\partial \Psi_2}{\partial R_\alpha}\right\rangle & \approx & \braket{\Psi_1(R_\alpha)}{\Psi_2(R_\alpha+\Delta R_\alpha)}/{\Delta R_\alpha}\nonumber\\
& = & \sum_{J=1}^6 \frac{C^*_{1J}(R_\alpha)C_{2J}(R_\alpha+\Delta R_\alpha)}{\Delta R_\alpha}.
\eea
Here, $R_\alpha$ is the $\alpha^{th}$ nuclear coordinate, $C_{jJ}$ is the coefficient in front of the $J^{th}$ configuration state function 
of the $j^{th}$ electronic state, and $\Delta R_\alpha$ is taken to be 0.0115 {\AA} for 
all nuclear coordinates.

To examine a role of the \gls{LBM} coordinate, 
two sets of models were generated: model $B$, including the \gls{LBM} coordinate, and model $N$, excluding it. Exclusion of \gls{LBM} is done by optimizing all stationary points at the \gls{CASSCF} level while constraining the nuclei of all lumiflavin's $\pi$-conjugated atoms in the lumiflavin plane. 
All out-of-plane coordinates of corresponding atoms have been fixed at their values found in minimizing 
the first excited state with respect to nuclear positions without constraints.
Hence, removing the single \gls{LBM} coordinate is done by removing a collection of fourteen rectilinear coordinates.

\subsection{Diabatic model}
\label{sec:diab-model}

Our model vibronic Hamiltonians are parametrized using the \gls{QVC} form~\cite{JahnTeller:2010} containing two electronic states
\bea\label{eq:QVC}
H & = &
  \begin{pmatrix} 
    H_D & V_c \\
    V_c & H_A
  \end{pmatrix},
\eea
where $H_D$ and $H_A$ are the individual vibronic Hamiltonians of $S_D$ and $S_A$, and $V_c$ is the vibronic coupling between the two electronic states:
\bea
H_k & = & -\frac{1}{2} \sum_\alpha^\Dim \frac{\partial_\alpha^2}{\partial x_\alpha^2} + V_k \hspace{0.35cm} \{k=D,A\}, \nonumber\\ \label{eq:Vk}
V_k & = & E_k + \sum_{\alpha}^\Dim v_{k,\alpha} x_\alpha + \sum_{\alpha,\beta}^\Dim x_\alpha K_{k,\alpha\beta} x_\beta \hspace{0.35cm} \{k=D,A,c\},\nonumber\\
\eea
$x_\alpha$ is the $\alpha^{th}$ mass-weighted nuclear coordinate, $E_{k}$, $v_{k,\alpha}$, and  $K_{k,\alpha\beta}$ are real-valued parameters used in the second order expansion of the model. The subscript $k$ takes values 
$D$, $A$, and $c$ to denote donor, acceptor, and coupling terms respectively. 
$2K_{k,\alpha\beta}$ represent the Hessian of the diabatic term $k$.
We use atomic units in \eq{eq:QVC} and throughout the paper. 
Boldface letters denote vectors and matrices quantities, for example, the matrix $K_{k,\alpha\beta}$ is written $\mat K_{k}$.
All models represent bound quantum systems which imposes some constraints on parametrization of the matrices $\mat K_k$ and is further elaborated in App.~\ref{app:param}. 
Any diabatic model can be transformed by a unitary transformation and will give the same dynamics. 
The \gls{DOF} associated with this unitary transformation are redundant and fixed by imposing a real 
parametrization as in \eq{eq:QVC} and by imposing $E_c=0$, which implies that diabatic and 
adiabatic states are matching at the coordinate origin. 

There are about 36,000 parameters for the full dimensional model $B$  and about 30,000 for model $N$. 
Therefore, usual fitting methods for 
adiabatic energies represented on a grid of points such as Levenberg-Marquardt~\cite{Levenberg:1944/qam/164,Marquardt:1963/jsiam/431} cannot be used here because the Jacobian and numerical gradients calculations in the parameter space are too cumbersome. Instead, we developed a two-step approach:

1) We generate a \gls{5D} \gls{QVC} model for a \gls{ps} of the most important modes.
The selected five \glspl{pm} correspond to the three vectors connecting the most important geometries of the molecular model (three minima of electronic states and the CI seam minumum) 
and the two branching space vectors at the minimum of the CI seam.
The corresponding Hamiltonian is $H_\ip$ and is parametrized accroding to \eq{eq:QVC}.
Parameters of $H_\ip$ are optimized in order to minimize the \gls{RMSD} of the adiabatic energies 
on a set of grid points along the \glspl{pm}. 
The optimization is done using the simplex algorithm.~\cite{Nelder:1965/cj/308} 
To suppress the appearance of negative curvatures in the diabatic states during the parameter optimization 
they have been assigned extra weights in a penalty function. 
For this optimization, the resulting models' \glspl{RMSD} are smaller than $2.4\cdot10^{-2}$ eV.

2) The remaining \glspl{sm} constitute a \gls{ss} with a Hamiltonian $H_\is$. They 
are included to build the full-dimensional model with a Hamiltonian 
\bea
H_\ifull & = & H_\ip + H_\is
\eea
where $H_\ip$ has been defined at the previous step and $H_\ifull$ is parametrized accroding to \eq{eq:QVC}.
$H_\is$ parameters are obtained using the steepest descent algorithm detailed in App.~\ref{app:steep}.
Resulting models' \glspl{RMSD} are smaller than $3\cdot10^{-3}$ eV Bohr$^{-1}$m$_e^{-1/2}$ 
for the fitting of the linear terms and smaller than $2.5\cdot10^{-4}$ eV Bohr$^{-2}$m$_e^{-1}$ 
for the fitting of the quadratic terms.
All directions corresponding to negative curvatures in the diabatic state Hessians of $H_\is$ have been
removed.
Using this two step procedure, two \gls{ps} models have been generated: the $B_\ip$ model  that 
includes the \gls{LBM} coordinate, and the $N_\ip$ model without the \gls{LBM} coordinate.
The corresponding full-dimensional models are labeled $B_\ifull$ and $N_\ifull$, and have 
dimensionalities of $\Dim=109$ and $\Dim=105$, respectively.

The electronic structure calculations for these fittings are done at the \gls{CASSCF} level.
The dynamical electron correlation is then included in the diabatic model by adjusting constant and linear terms of diabatic Hamiltonians to match 
the \gls{XMCQDPT2} energies on a set of grid points connecting the stationary points. 
This step is motivated by the fact that the quasi-diabatic states correspond to a dominant configuration or 
a set of configurations represented by a particular element of the diabatic Hamiltonian, 
on which the \gls{XMCQDPT2} correction will have a uniform effect for all nuclear geometries.
This correction can be done only for \gls{ps} models because such a grid cannot be built for the full dimensional space.
This correction was applied on model $B_\ip$.

\subsection{Lifetime calculation}

The electron transfer dynamics is followed as the population of the the donor state with the \gls{NFGR} method.~\cite{Izmaylov:2011/jcp/234106,Endicott:2014/jcp/034104}
The population of the donor state is expressed as
\bea\nonumber
P_D(t) & = & \braOket{D}{\tr_B\{\rho(t)\}}{D}\\\label{eq:pop}
& = & \braOket{D}{\tr_B\{\Lambda(t)\rho(0)\Lambda(t)^\dagger\}}{D},
\eea
where $\tr_B\{\dots\}$ is the trace over all model nuclear \gls{DOF}, $\ket{D}$ is the donor electronic state, $\rho(t)$ is the total density matrix of the quantum system at time $t$, and $\Lambda(t)$ is the unitary evolution operator. Under the assumption that the coupling element $V_c$ is small, $\Lambda(t)$ can be expanded as a perturbation series of $V_c$. In the second order cumulant expansion the donor population becomes
\bea\label{eq:NFGR}
P_D(t) & = & e^{-\int_0^t dt' \int_0^{t'} dt'' f(t',t'')},
\eea
where $t'$ and $t''$ are time integration variables and the integrand is a correlation function
\bea\label{eq:f}
f(t',t'') & = & 2\Re\left\langle V_c(t') V_c(t'') \right\rangle_T \nonumber\\
 & \hspace{-2cm} = & \hspace{-1cm} 2\Re\frac{\tr\{e^{iH_Dt'} V_c e^{-iH_A (t'-t'')} V_c e^{-iH_D t''} e^{-H_G/k_BT}\}}{\tr\{e^{-H_G/k_BT}\}},\nonumber\\
\eea
where the trace is over nuclear \gls{DOF} and $k_B$ is the Boltzmann constant. $\rho(0)$ is chosen as a Boltzmann distribution at temperature $T$ of some ground state with Hamiltonian $H_G$ (to be defined later) placed in the donor state. Such placement is justified assuming an ultra-fast excitation pulse. 
$f(t',t'')$ is an analytic function of diabatic Hamiltonian parameters and temperature,  its explicit expression 
is given in the appendix of Ref.~\citenum{Endicott:2014/jcp/034104}.
The average lifetime of the system in the donor state, $\tau_D$, is calculated as 
\bea\label{eq:lifetime}
\tau_D & = & \frac{\int_0^\infty t P_D(t) dt}{\int_0^\infty P_D(t) dt}.
\eea
Parameter optimizations for both models (N and B) and 
quantum dynamics calculations were done using the Matlab program.~\cite{Matlab:2012b}

\section{Results}
\label{sec:results}

\subsection{Electronic structure calculations and \protect\gls{PES} exploration}

Four stationary points were optimized: 1) the minimum of the ground electronic state,  $M_0$; 2) the $S_1$ minimum 
with the excited electron localized on lumiflavin, $M_D$ ; 3) the $S_1$ minimum with the excited electron localized 
on the \gls{CPD} lesion, $M_A$; and 4) the minimum of the \gls{CI} seam between $S_1$ and $S_2$, $M_{CI}$.
Geometries of these configurations and the non-adiabatic coupling vector calculated at $M_{CI}$ 
can be found in the supplemental materials.~\cite{SuppInfo}
We give here an approximate energy profile, 
obtained by single point calculations at the \gls{CASSCF}/6-31G$^*$ level of theory, that interpolates the optimized stationary points in Fig.~\ref{fig:profile}-a.
\gls{CASSCF} energies corrected by \gls{XMCQDPT2} are given in Fig.~\ref{fig:profile}-b. The electronic excitation energies at $M_0$ reproduce the energies calculated in Ref.~\citenum{Domratcheva:2011/jacs/18172}, which validates the molecular model.
As it can be noticed on Fig.~\ref{fig:profile}-b the \gls{XMCQDPT2} corrections shift all states in energy 
and their minima positions in configuration space: $M_D$ is not a local minimum in $S_1$, 
and the minimum of the \gls{CI} seam is displaced.
The corrected \gls{CI} seam minimum is closer to $M_D$. Thus, the energy difference between $M_{CI}$ and $M_D$ 
is expected to be smaller than that at the \gls{CASSCF} level of theory. 
It can be anticipated that the \gls{ET} dynamics is faster for \gls{XMCQDPT2} 
corrected models than for \gls{CASSCF} models.
\begin{figure}
  \begin{tabular}{cc}
    \raisebox{4cm}{a)}&\includegraphics[width=0.45\textwidth]{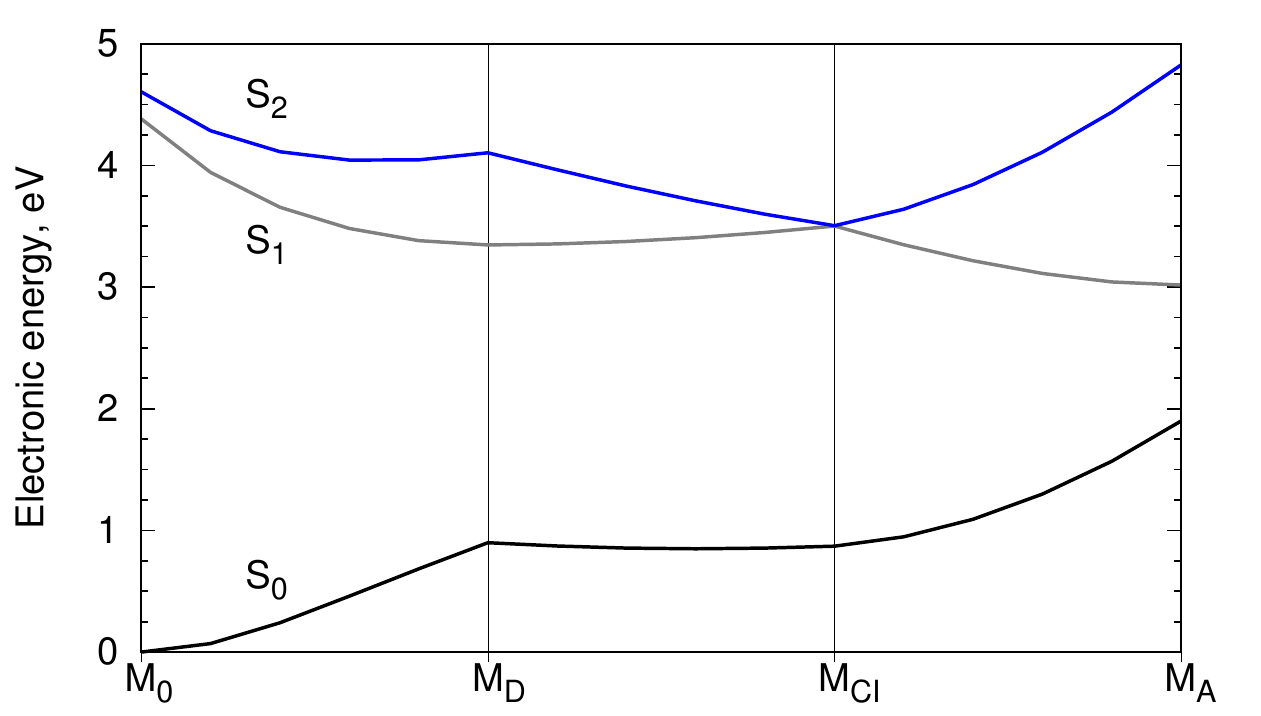}\\
    \raisebox{4cm}{b)}&\includegraphics[width=0.45\textwidth]{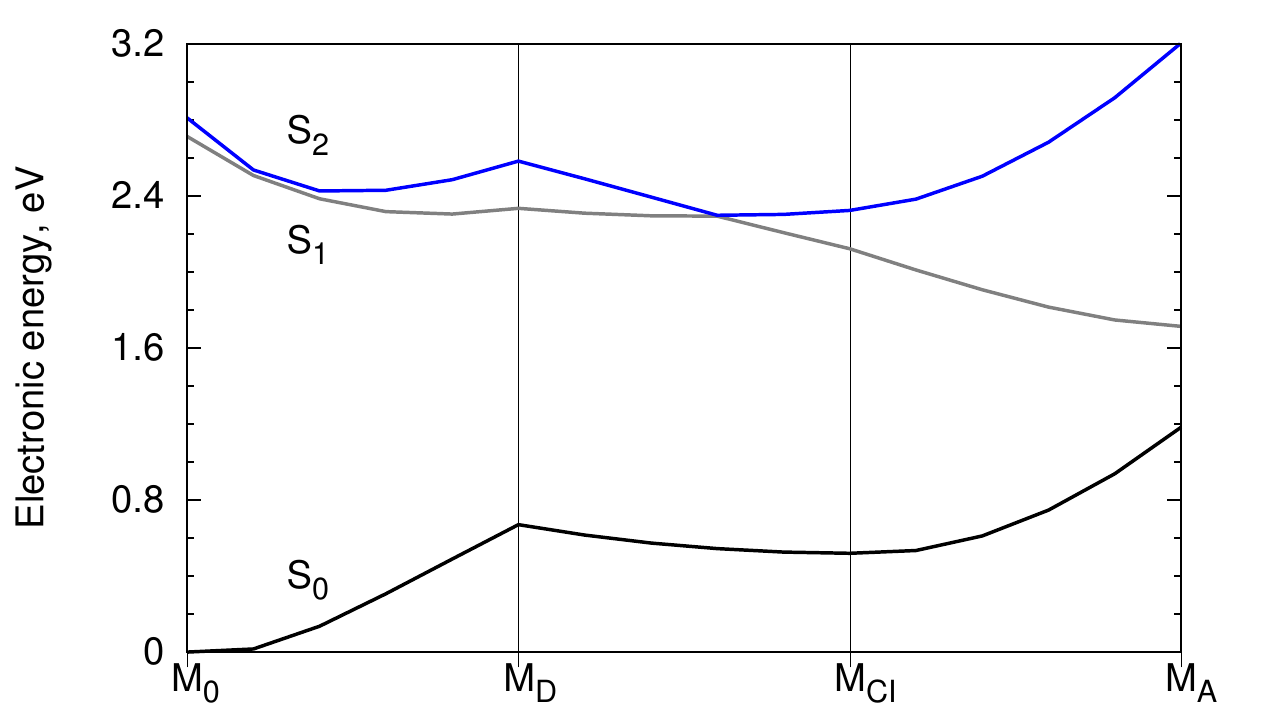}
  \end{tabular}
  \caption{
a) Approximate profile at \protect\gls{CASSCF}/6-31G$^*$ level of theory. The curves are obtained by rectilinear interpolations between two consecutive stationary geometries.
b) Approximate profile at \protect\gls{XMCQDPT2} level of theory. The curves are obtained by rectilinear interpolations between two consecutive stationary geometries obtained at \protect\gls{CASSCF}/6-31G$^*$ level of theory.}
  \label{fig:profile}
\end{figure}

We observed an exchange of configuration in the \gls{CASSCF} electronic states on each side of the \gls{CI} and the quasi-diabatic states $S_D$ and $S_A$ (defined in Sec.~\ref{sec:MolMod-elec}) are such that: $S_1\equiv S_D$ and $S_2\equiv S_A$ at $M_D$, while $S_1\equiv S_A$ and $S_2\equiv S_D$ at $M_D$. This observation supports our approach to building a two-state crossing diabatic model for this \gls{ET} reaction. The conical intersection seam minimum and adiabatic surfaces around it are plotted along the branching plane modes in Fig.~\ref{fig:CI}. 
Figure~\ref{fig:CI} shows a very small splitting along one of the directions. Therefore, the crossing subspace is 
almost $\Dim-1$ dimensional, where $\Dim$ is the total number of nuclear dimensions, which
is consistent with small diabatic coupling between the crossing 
diabats.~\cite{Fernandez:2000/jacs/7528,Blancafort:2001/JACS/722}
\begin{figure}
  \begin{tabular}{c}
    \includegraphics[width=0.5\textwidth]{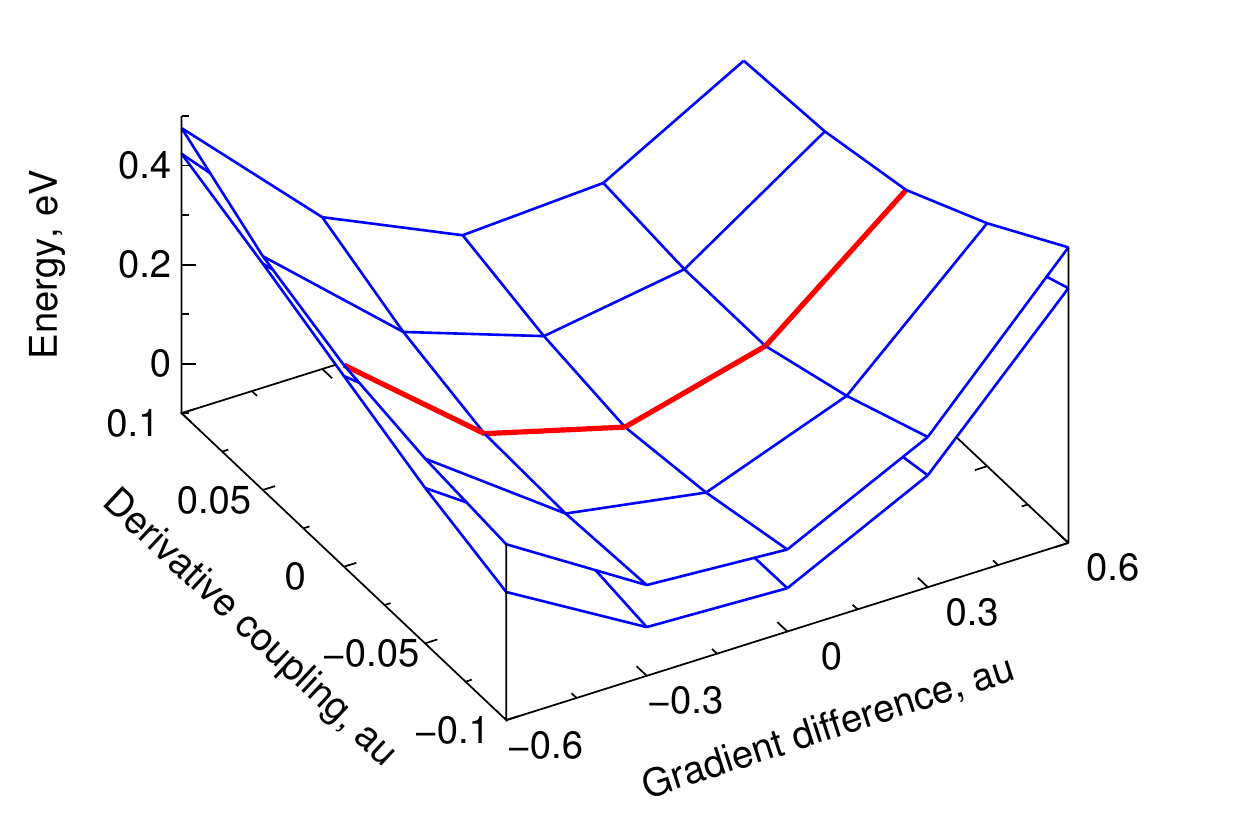}\\
  \end{tabular}
  \caption{Adiabatic electronic energies in the branching space in the vicinity of the minimum of the \protect\gls{CI} seam
  obtained at the \protect\gls{CASSCF}/6-31G$^*$ level. The coordinates' origin is set at $M_{CI}$. The two directions and their scales are defined by orthonormalizing the covariant gradient difference 
  with respect to the non-adiabatic coupling vector in mass-weighted coordinates. 
  The lower surface is the first excited state, $S_1$, while the upper surface is the second 
  excited state, $S_2$. The red line indicates a line of quasi-degeneracy between the two states.}
  \label{fig:CI}
\end{figure}

A local electronic excitation of the lumiflavin leads to 
reorganization of the $\pi$ orbital system that exerts a force on the nuclei in the Franck-Condon 
region (see the gradient of $S_1$ energy at $M_0$ geometry in Fig.~\ref{fig:vectors}-a). 
Moving from the Franck-Condon region along the gradient 
modifies bond orders between nuclei in the lumiflavin plane. Therefore, it also changes the hybridization of 
nitrogens responsible for the bent conformation of lumiflavin from $sp^3$ to a $sp^2$ and leads to 
planar lumiflavin in the excited electronic state. 
The relative coupling between the two directions in the $S_1$ Hessian is $2.16$,~\cite{misc:coupling}
 which indeed proves that the force at the \gls{FC} point will induce a change of the lumiflavin conformation from bent to planar. 
This analysis agrees with the planar geometry of lumiflavin found at $M_D$. Hence, main nuclear coordinates involved in motion from $M_0$ to $M_D$ are lumiflavin out-of-plane coordinates 
as can be seen from the corresponding displacement vector given in Fig.~\ref{fig:vectors}-c.
\begin{figure}
  \begin{tabular}{|cc|cc|}\hline
    \raisebox{2.4cm}{a)}&\includegraphics[width=0.20\textwidth]{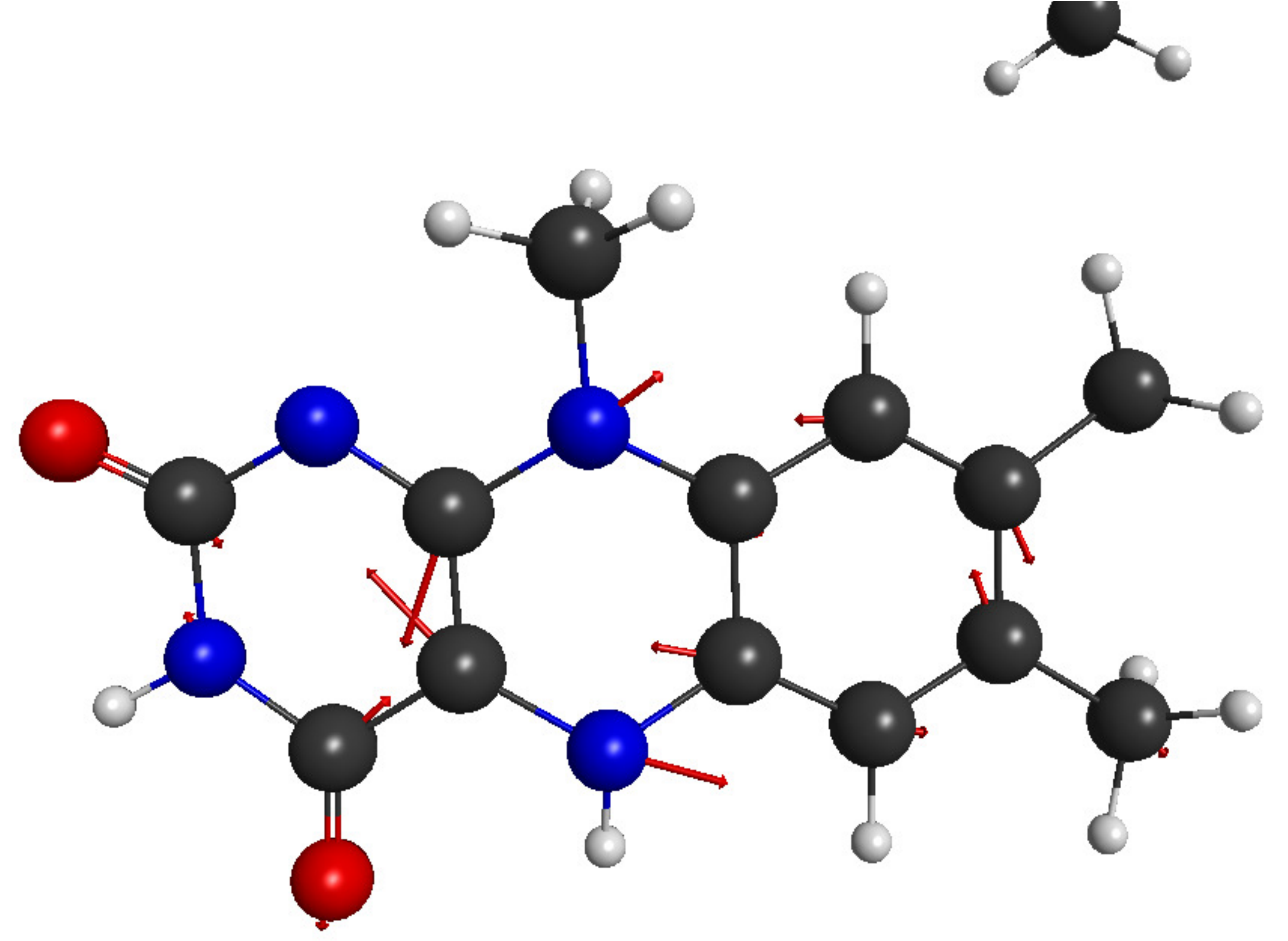}                        & \raisebox{2.4cm}{b)}&\includegraphics[width=0.20\textwidth]{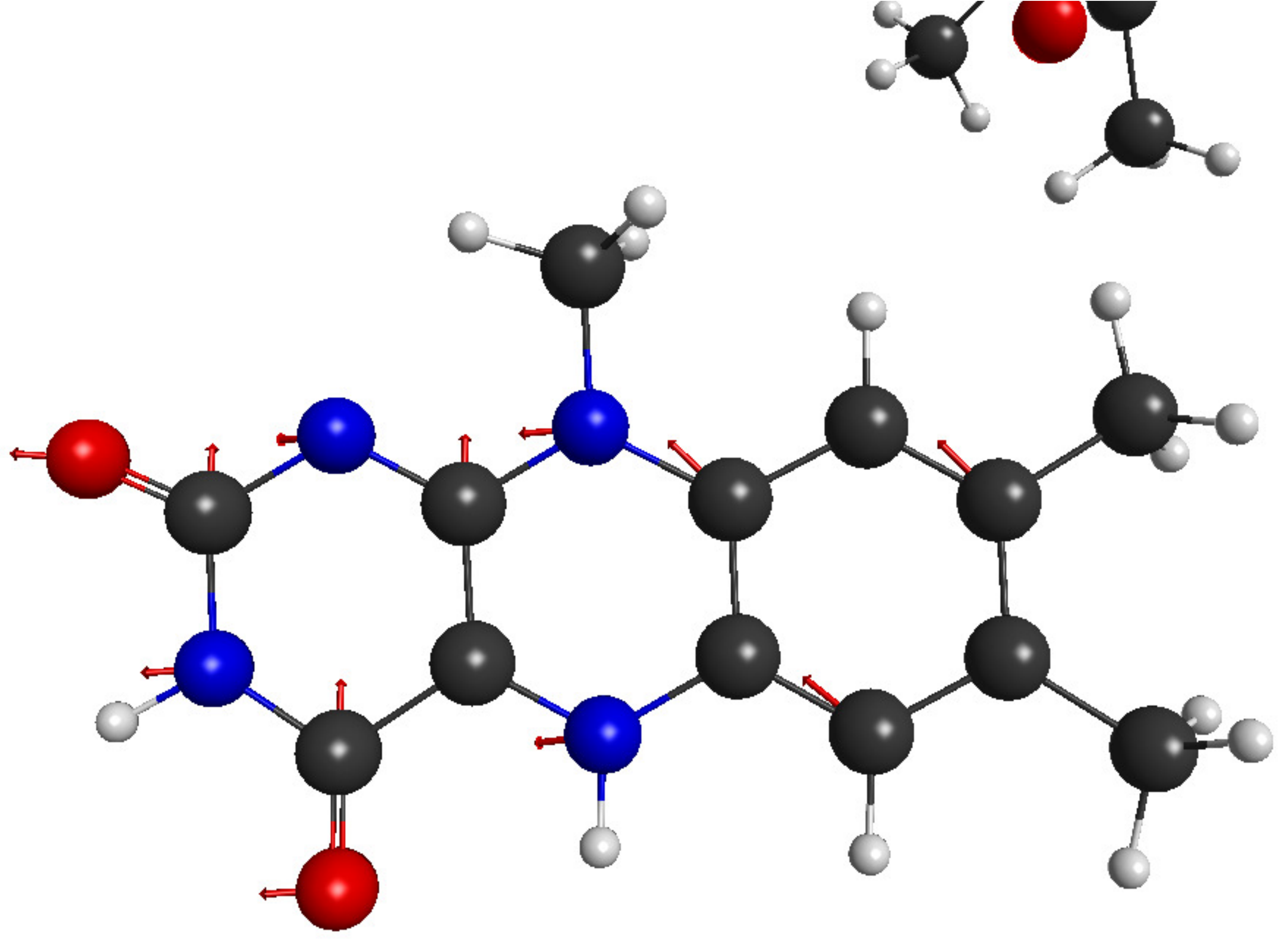}\\\hline
    \raisebox{1.2cm}{c)}&\includegraphics[trim=0 0 250 200,clip,width=0.20\textwidth]{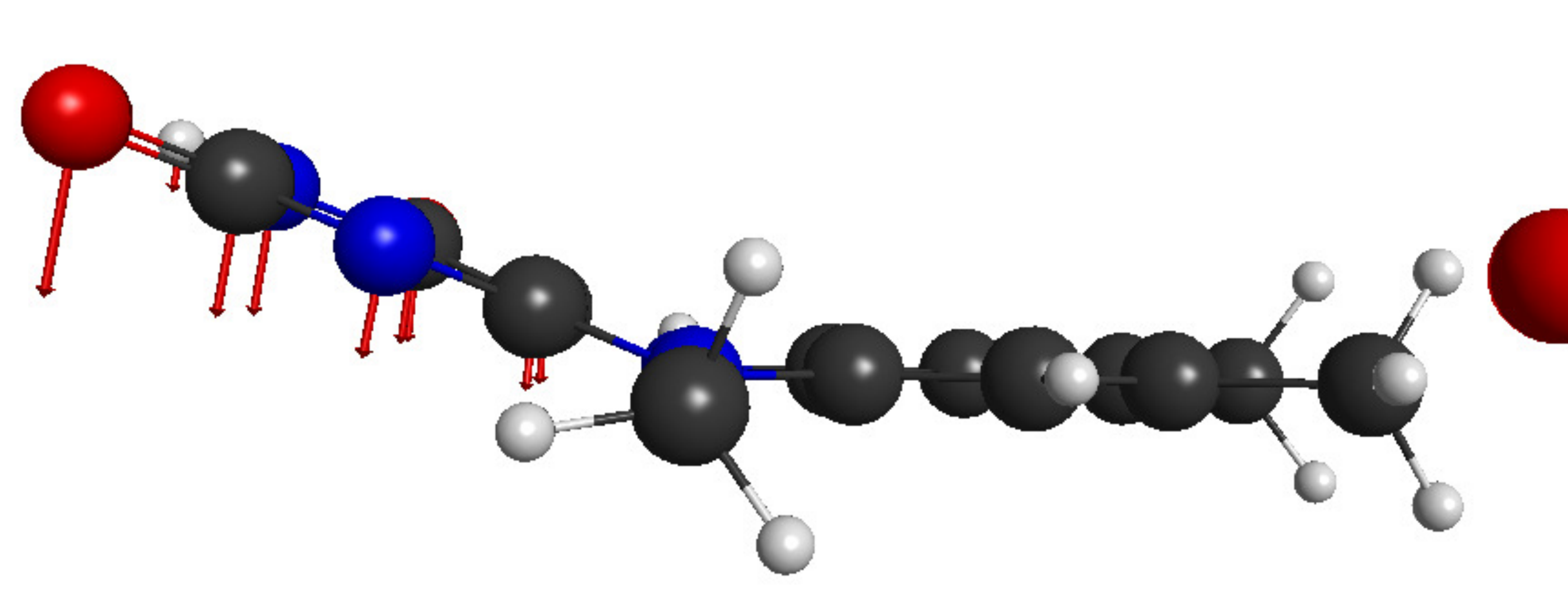} & \raisebox{1.2cm}{d)}&\includegraphics[trim=200 140 425 300,clip,width=0.20\textwidth]{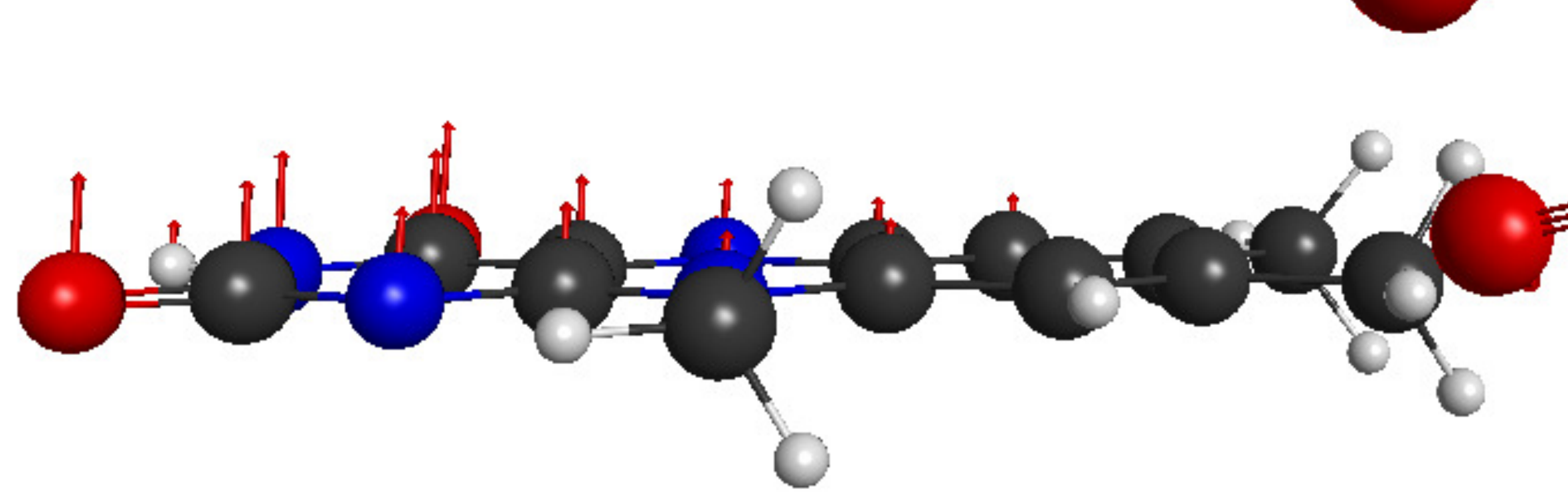}\\\hline
    \raisebox{2.8cm}{e)}&\includegraphics[width=0.17\textwidth]{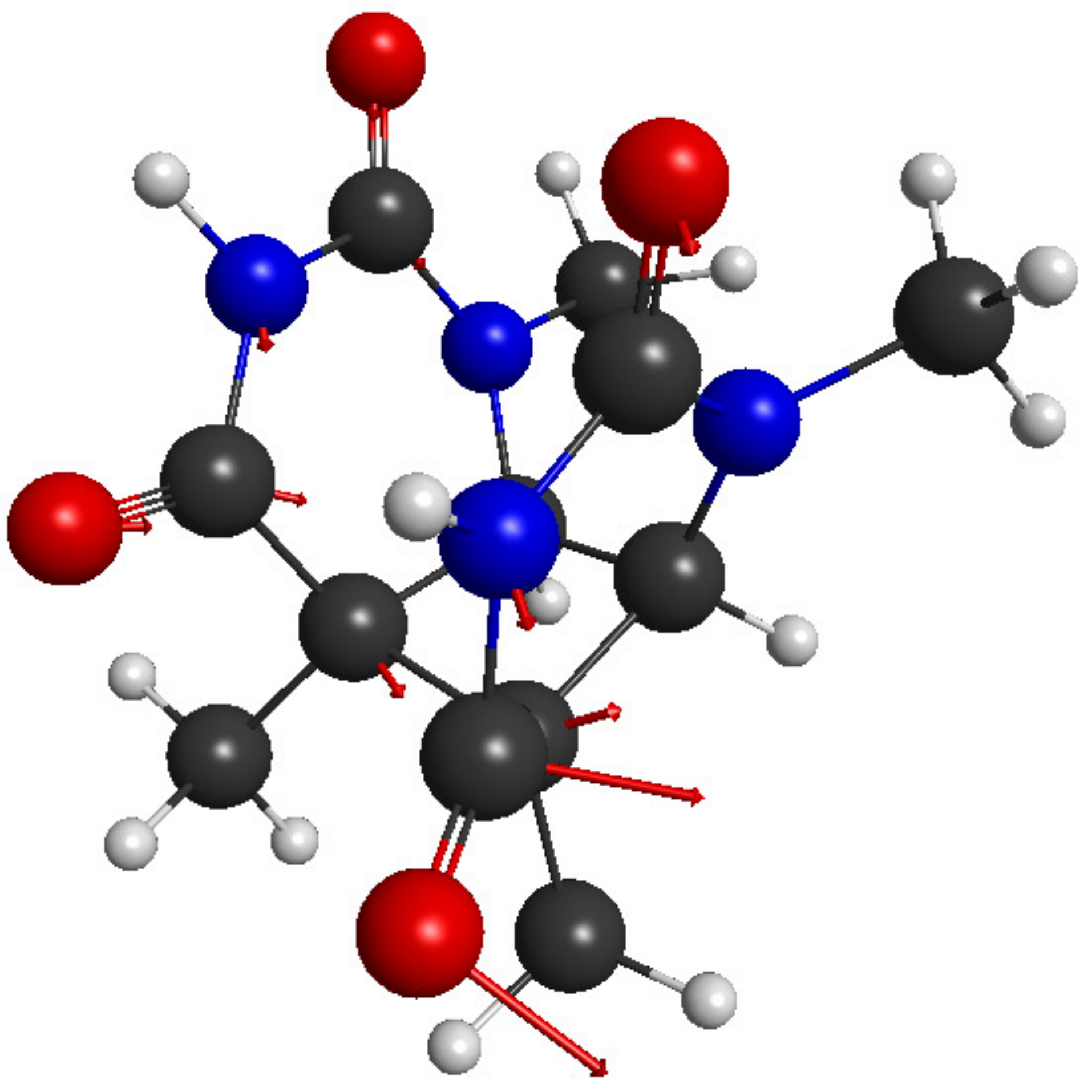}                        & \raisebox{2.8cm}{f)}&\includegraphics[trim=300 0 300 50,clip,width=0.20\textwidth]{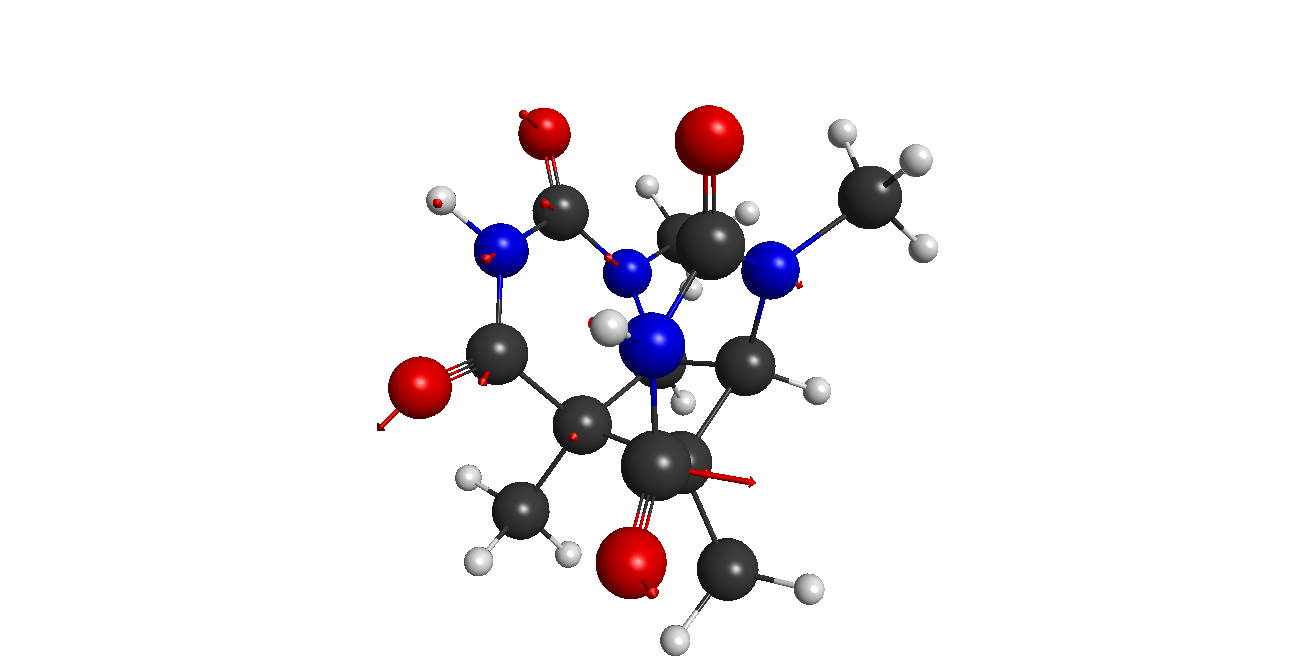}\\\hline
  \end{tabular}
  \caption{
a) Energy gradient vector at $M_0$ geometry on the first excited electronic state.
b) Derivative coupling for the \protect\gls{CI} between first and second electronic excited states at $M_{CI}$.
c) Displacement vector from geometry $M_0$ to $M_D$. This vector is mainly the out-of-plane motion of the lumiflavin nuclei.
d) Displacement vector from $M_D$ to the \protect\gls{CI}.
e) Gradient difference for the \protect\gls{CI} between first and second electronic excited states at $M_{CI}$.
f) Displacement vector from $M_D$ to $M_A$.}
  \label{fig:vectors}
\end{figure}

The \gls{ET} process proceeds from the flavin $\pi^*$ system to the $\pi^*$ orbital of the \gls{CPD} carbonyl group. 
Thus, the carbon of the carbonyl group on \gls{CPD} changes its hybridization from $sp^2$ to $sp^3$ and
 the carbonyl orientation with respect to the thymine ring becomes bent, while changes in the lumiflavin $\pi$ system cause 
 bond reordering. Both motions are described by the branching space vectors 
 given in Figs.~\ref{fig:vectors}-b and~\ref{fig:vectors}-e.

The \gls{LBM} motion was also found to participate in accessing the $M_{CI}$ from $M_D$ or $M_A$ as can be seen in Fig.~\ref{fig:vectors}-d. However, these displacement vectors have small amplitudes 
and removing the \gls{LBM} coordinate must not change the $M_{CI}$ position significantly.
Reoptimization of the \gls{CI} seam minimum 
without including the \gls{LBM} coordinate leads to geometry that is very similar to
and only 30 meV higher in energy than that with including the \gls{LBM} coordinate.
Stationary point optimizations with constrained \gls{LBM} for the model $N$ gave new geometries 
that can be found in the supplemental materials.~\cite{SuppInfo}
Only the minimum of the ground state $M_0$ is significantly affected while other stationary points 
were already almost planar without the constraints, and their changes in energy are of only few dozens of meV.
The \gls{CI} seam still shows a $\Dim-1$ crossing subspace, which implies a very small coupling 
between the electronic states. However, \gls{LBM} participates in connecting the excited states 
and the CI seam minima. 
Therefore, the distances between $M_{CI}$, $M_D$, and $M_A$ are smaller after 
removing the \gls{LBM} coordinates.

\subsection{\protect\gls{ET} lifetimes}

The five \glspl{pm} connect the stationary points ($M_0$, $M_D$, $M_A$, and $M_{CI}$) and describe the two branching space vectors at $M_{CI}$. 
A grid is built along these five dimensions at the \gls{CASSCF} level for the fitting procedure. 
A visual comparison of the errors between the \gls{CASSCF}/6-31G$^*$ calculations and the $B_\ip$ model 
along the approximate profile and along the branching space modes at the vicinity of $M_{CI}$
can be found in the supplemental materials.~\cite{SuppInfo} 

Initial conditions for the quantum dynamics are chosen as a Boltzmann distribution in $S_D$ at $298$ K. 
This choice is made assuming that the relaxation of the nuclear subsystem in $S_D$ does not affect
appreciably the \gls{ET} process. 
This assumption is based on approximate orthogonality 
between the relaxation direction $(M_0,M_D)$ and 
active coordinates of the \gls{ET} process, which are given by the direction $(M_D,M_A)$ 
and the branching space vectors.
Indeed, $(M_0,M_D)$ is almost orthogonal to the direction $(M_D,M_A)$ with an angle 
$\widehat{M_0 M_D M_A}=76^\circ$ (see Figs.~\ref{fig:vectors}-a and ~\ref{fig:vectors}-f) 
and the two directions are very weakly coupled with a relative coupling of $9.5\cdot10^{-3}$ in the Hessian.~\cite{misc:coupling}
Moreover, the direction $(M_0,M_D)$ is orthogonal to the branching space (see Figs.~\ref{fig:vectors}-b and~\ref{fig:vectors}-e) with an angle of $87^\circ$ so the crossing region is not easily accessed by the system from $M_0$.
Long \gls{ET} time-scale provides the system with enough time to relax on the excited state before the \gls{ET} process.
For a Boltzmann distribution in $S_D$ as the initial condition, the correlation function [\eq{eq:f}]
depends only on the difference $t'-t''$. Thus $f(t',t'')=f(0,t''-t')\equiv f(t''-t')$ with
\bea\label{eq:fDelta}
f(t) & = & 2\Re\left\langle V_c(0) V_c(t) \right\rangle_T.
\eea
With this simplification, the donor population \eq{eq:NFGR} takes a simpler form
\bea\label{eq:simpleP}
P_D(t) & = & e^{\int_0^{t} dt' f(t') t'} e^{-t\int_0^{t} dt' f(t')}.
\eea

The model $B_\ip$ shows an exponential decay with a full population transfer 
(red curve in Fig.~\ref{fig:pop-5D-CAS}) due to vibrational state resonances between 
$S_D$ and $S_A$.~\cite{Endicott:2014/jcp/034104}
The theoretical lifetime is $10$ ps, which has to be compared with the experimental one of $170$ ps.~\cite{Kao:2005/pnasusa/16128}
Given the simplified assumptions used to construct our model, we consider this to be in qualitative agreement. 
The long time scale indicates that the $B_\ip$ model correctly accounts for the small coupling between the electronic states.
The energies from single points calculation at the \gls{XMCQDPT2} level of theory given in Fig.~\ref{fig:profile}-b are used to adjust the diabatic elements of the model $B_\ip$. As expected from the displacement of the \gls{CI} seam minimum, 
 after adding the \gls{XMCQDPT2} corrections, the lifetime becomes shorter, $2$ ps.
\begin{figure}
  \begin{tabular}{c}
    \includegraphics[width=0.45\textwidth]{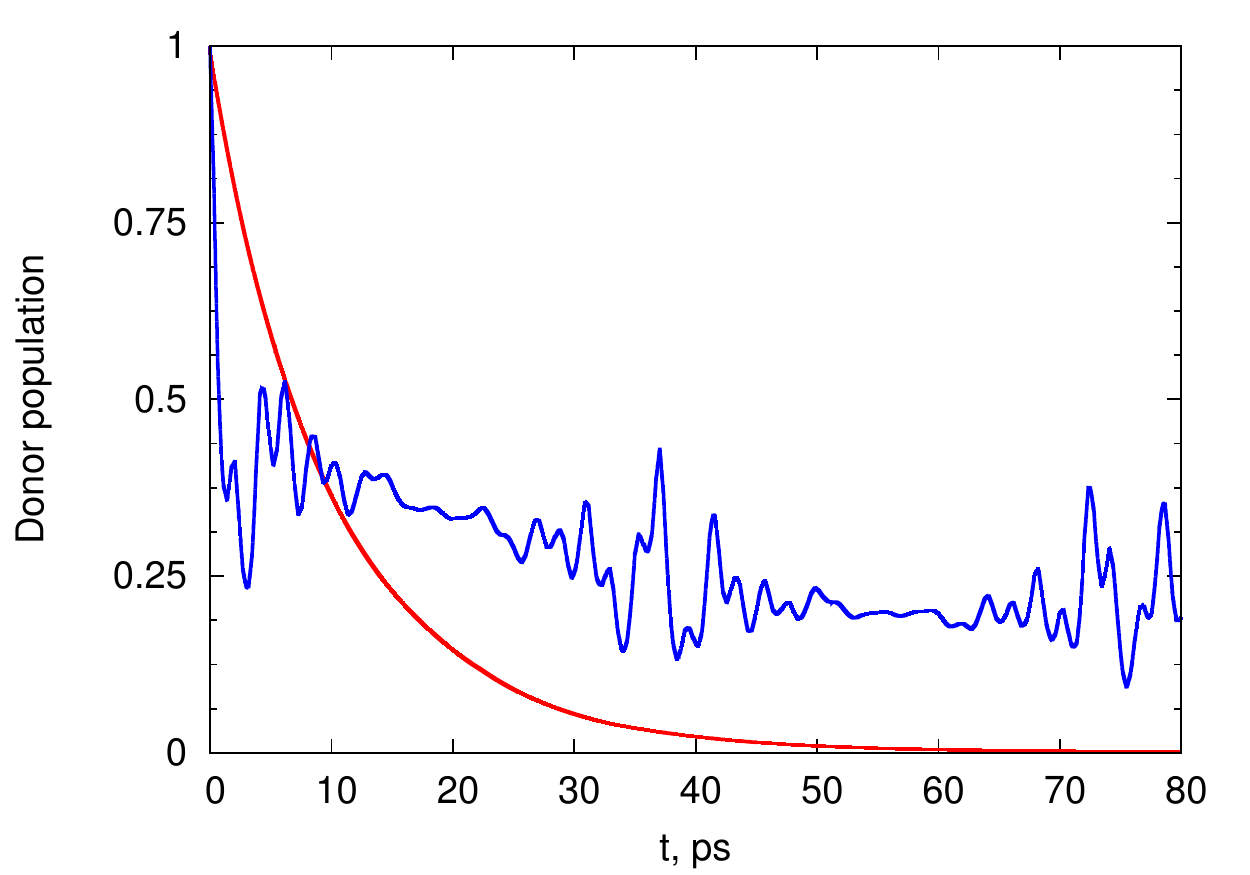}\\
  \end{tabular}
  \caption{Time evolution of the $S_D$ population for the \protect\gls{ps} model including ($B_\ip$ in red) or not ($N_\ip$ in blue) the lumiflavin butterfly motion.
           Vibronic resonances in $B_\ip$  induce a full population transfer while the absence of such resonances in $N_\ip$ causes oscillations.}
  \label{fig:pop-5D-CAS}
\end{figure}

\subsection{Variations of the \protect\gls{ps} model}
\label{sec:subsyst}

The lifetime is sensitive to parameter variations in quadratic terms. 
These terms can be separated in three groups:
1) quadratic terms in $V_c$ (diabatic coupling) contained in the matrix $K_{c,\alpha\beta}$ used in \eq{eq:QVC};
2) contributing to differences between the donor and acceptor state normal mode frequencies $\{\Omega_{D,\alpha}\}$ 
and $\{\Omega_{A,\alpha}\}$; 
3) contributing to Duschinsky rotation, a transformation $\{R_{\alpha\beta}\}$ to obtain the acceptor state 
normal modes $Q_{A,\alpha}=\sum_{\beta}R_{\alpha\beta}Q_{D,\beta}$ from the donor 
state normal modes $\{Q_{D,\alpha}\}$.
The last two groups can be linked to coefficients $K_{D,\alpha\beta}$ and $K_{A,\alpha\beta}$ used in \eq{eq:QVC}:
\bea\label{eq:diag-Qterms}
\sum_{\alpha,\beta} x_\alpha K_{D,\alpha\beta} x_\beta & = & \sum_{\alpha} \frac{1}{2}\Omega_{D,\alpha}^2 Q_{D,\alpha}^2, \nonumber\\
\sum_{\alpha,\beta} x_\alpha K_{A,\alpha\beta} x_\beta & = & \sum_{\alpha} \frac{1}{2}\Omega_{A,\alpha}^2 Q_{A,\alpha}^2 \nonumber\\
& = & \sum_{\alpha} \frac{1}{2}\Omega_{A,\alpha}^2 \Big[\sum_{\beta}R_{\alpha\beta}Q_{D,\beta}\Big]^2,
\eea
and it can be noticed that $\mat K_{A}=\mat K_{D}$ if $\mat\Omega_{A}=\mat\Omega_{D}$ 
and $\mat R=\mat 1$. To see effects of these terms on lifetimes we contrasted 
the results for full QVC models with those obtained omitting each group of terms 
separately (see Table~\ref{tab:lifetimes}).
The model's variations modify the lifetime such that the \gls{ET} process 
can be blocked (a lifetime larger by more than four orders of magnitude), 
which shows the importance of the quadratic terms in the model. 
\begin{table}
  \centering
    \begin{tabular}{@{\hspace{0.4cm}}c@{\hspace{0.4cm}}c@{\hspace{0.4cm}}c@{\hspace{0.4cm}}c@{\hspace{0.4cm}}c}\hline\hline
  model   &      QVC    &  $\mat K_c$=0   & $\mat R=\mat 1$ & $\mat\Omega_{A}=\mat\Omega_{D}$ \\\hline
 $B_\ip$  &     $10$    &     $10$        &      $11$       &              NT                 \\
$B_\ifull$&      $1$    &      $2$        &        NT       &              NT                 \\
$N_\ifull$&     $0.3$   &     $0.5$       &        NT       &               $4$                \\\hline\hline
    \end{tabular}\vspace{0.15cm}
  \caption{Lifetimes given in picoseconds for different variations of the diabatic models including. NT: no transfer happens with an estimate of the lifetime larger than 0.1 $\mu$sec ($\tau>10^{5}$).}
  \label{tab:lifetimes}
\end{table}
Removing the quadratic terms in $V_c$ ($\mat K_{c}=0$) does not affect strongly the lifetime, 
which can be understood because the coupling between the two states is already very small.
In contrast, imposing the same quadratic terms in both states results in much longer lifetimes. 
After decomposing $\mat K_D-\mat K_A$ in terms of Duschinsky rotation and frequency differences, 
it is easy to realize from Table~\ref{tab:lifetimes} that imposing the same vibrational frequencies 
for both diabatic states changes 
lifetimes drastically by modifying resonances between the two electronic states.

\subsection{Full dimensional models}
\label{sec:bath}

Adding residual modes makes the correlation functions \eq{eq:fDelta} decay within 
few tens of femtoseconds (see Fig.~\ref{fig:correl}).
\begin{figure}
    \includegraphics[width=0.45\textwidth]{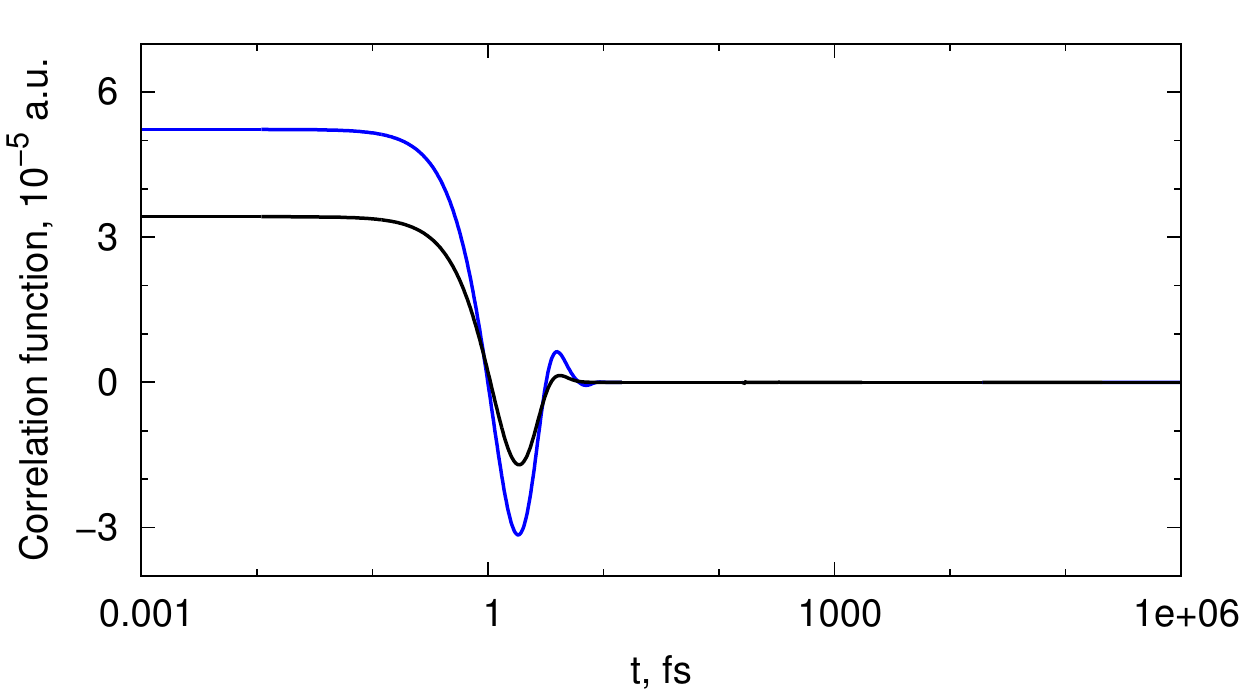}
  \caption{Correlation function in logarithmic time axis for the various full-dimensional models. Blue lines are for models that include the \protect\gls{LBM}, while black lines does not.}
  \label{fig:correl}
\end{figure}
Defining $t_0$ as the decay time in which the correlation function, $f(t)$, becomes numerical 0. 
For both models with residual modes $t_0\approx 20$ fs, hence $t_0\ll\tau_D$ (ps) and we can make 
a Markovian approximation for the donor population given by \eq{eq:simpleP} 
\bea\label{eq:MarkovP}
P_D(t) & \approx & e^{\int_0^{t_0} dt' f(t') t'} e^{-t\int_0^{t_0} dt' f(t')}.
\eea
According to \eq{eq:lifetime} and \eq{eq:MarkovP} the donor state lifetime $\tau_D$ is defined as
\bea\label{eq:Markovlifetime}
\tau_D & = & \left[\int_0^{t_0} dt' f(t')\right]^{-1}.
\eea

Including additional modes generally increases the reorganization energy, and according to the Marcus theory 
is expected to slow down the \gls{ET} process.
In contrast, in our simulations, adding the \glspl{sm} decrease the lifetime (see $B_\ifull$ in Table~\ref{tab:lifetimes}).
 To understand this effect, one has to remember that the additional 
modes also contribute to the coupling term,
$V_c$. Even if the individual contributions to the linear coupling vector $\mat v_c$ (\eq{eq:Vk}) from each \gls{sm} is small, 
owing to the overall large number of these modes, 
their total contribution is of the same order of magnitude as the contribution from the \gls{ps}. 
Their effect can be estimated by looking at the ratio of the coupling vector norm, $||\mat v_c||$, over the norm of the difference of the linear term for diabatic energies, $||\mat v_A-\mat v_D||$, the latter is associated with the reorganization energy. For $B_\ip$ this ratio is $0.3$, while for $B_\ifull$ it is $0.9$.
The inclusion of \glspl{sm} emphasizes the importance of the Duschinsky rotation as can be seen in Table~\ref{tab:lifetimes}. Therefore, it is crucial for adequate modelling of $B_\ifull$ to account for different 
orientations of normal modes in the two electronic states.

\subsection{Role of the \protect\gls{LBM} motion}
\label{sec:noLBM}

The procedure to obtain the parameters for the model $N_\ip$ is the same as for the model $B_\ip$ except that the \gls{LBM} motion is constrained for geometry optimizations as described in Sec.\ref{sec:MolMod-elec} and the new stationary points are used in the fitting procedure. 
The same number of stationary points are obtained for $N_\ip$ as for $B_\ip$. Hence, both models are \gls{5D}.
Population dynamics in these models (Fig.~\ref{fig:pop-5D-CAS}) show that the \gls{LBM} motion
has a strong impact on the dynamics.
In contrast to the dynamics in $B_\ip$, the $N_\ip$ dynamics does not show a full transfer but rather 
has population oscillations. Moreover, the latter dynamics starts with a fast initial drop within the first hundred  femtoseconds followed by oscillations.
 The oscillations appearing for the $N_\ip$ model indicate that the vibrational states of the 
donor and acceptor wells are not in resonance. Including the low frequency \gls{LBM} mode
in the $B_\ip$ model increases the density of vibrational states and induces resonances, 
which remove population oscillations and lead to a complete population transfer. 
Thus, one role of the \gls{LBM} motion is to induce decoherence.
Because of the oscillations in $N_\ip$, no lifetime can be extracted in this case.
Adding the \glspl{sm} causes decoherence and complete \gls{ET}. 
Therefore, we can calculate the lifetime for $N_\ifull$ and compare it with the lifetime obtained in $B_\ifull$.
The \gls{LBM} motion slows down the \gls{ET} by almost one order of magnitude in full-dimensional models (see $N_\ifull$ compared to $B_\ifull$ in Table~\ref{tab:lifetimes}).
The slower \gls{ET} in $B_\ifull$ does not happen because of a smaller coupling ($||\mat v_c||$ is of the same magnitude for $B_\ip$ and $N_\ip$), it is rather due to smaller Franck-Condon factors appearing from the Duschinsky 
rotation terms as can be seen in Table~\ref{tab:lifetimes}. Therefore, another role of the \gls{LBM} motion 
is to decrease Franck-Condon factors.

\section{Conclusion}
\label{sec:conclusion}


Above, we present the first diabatic \gls{PES} models for the \gls{ET} process that initiates the \gls{CPD} lesion repair in the CPD-photolyase complex.
Our models give lifetimes between one (for the $109$-dimensional model) and ten picoseconds (for the $5$-dimensional model), which is
one order of magnitude faster (but still relatively close if compared with other ultrafast processes) than the 
\gls{ET} timescale measured experimentally (170 ps). 
Slow rates of \gls{ET} in our models is attributed to small diabatic coupling between the two diabatic 
electronic states, and even if the crossing region between electronic states is easily 
accessible, the system remains diabatically trapped.
We showed that the quadratic terms used in the expansions of the diagonal elements of our diabatic model (frequency differences and Duschinsky rotation of normal modes) must be included as they are essential 
for modelling Franck-Condon factors which regulate the \gls{ET} process.
In contrast, the quadratic terms in the off diagonal diabatic couplings are negligible. 
Investigation concerning the inclusion or exclusion of the \gls{LBM} coordinate showed 
the importance of this mode for dynamics as it enhances incoherent transfer 
by increasing density of vibrational states due to its low vibrational frequency. 
The intrinsic properties of the donor-acceptor complex at the geometry controlled 
by binding to the protein are favorable for \gls{ET} without a ``mediator'' as proposed 
for adenine.~\cite{Liu:2012/jacs/8104} However, adenine might be involved in ``adjusting" 
the donor and acceptor energies in the protein DNA complex 
(as opposed to the considered here CPD-flavin complex), 
where for instance, negative phosphates in the proximity of donor and 
acceptor may alter the state energies.~\cite{Moughal:2013/jctc/4644}


In future work we will focus on the effects of amino acid residues present 
in the active site and will assess the importance of the protein environment in 
order to obtain quantitative estimations of the lifetime. 
A simple method we developed in this paper can be also applied to other steps of the \gls{CPD} repair process, 
in particular the electron back transfer. It can also be applied to the \acrlong{64PP} repair process and give insights on the futile electron back transfer~\cite{Li:2010/Nature/887} by comparison with the \gls{CPD} process.
Models generated in this paper can be further used to study the effect of exciting the system 
with incoherent sunlight~\cite{Tscherbul:2014/JPCA/3100,Joubert:2015/jcp/134107} 
rather than coherent laser light used in transient spectroscopy experiments.~\cite{Kao:2005/pnasusa/16128}

\section{Aknowledgments}
LJD thanks Dvira Segal for providing some computational resources and the European Union Seventh Framework Programme (FP7/2007-2013) for financial support under grant agreement PIOF-GA-2012-332233.
TD acknowledges support  through the Minerva program of the Max Planck Society.

\appendix

\section{Diabatic potential parametrization}
\label{app:param}

The Hessians of the diabatic states, $\mat K_D$ and $\mat K_A$, in \eq{eq:QVC} 
must be real symmetric positive definite matrices. Hence, we naturally choose to 
parameterize $\mat K_D$ and $\mat K_A$ using the Cholesky decomposition
\bea\label{eq:K-param}
\mat K_k & = & \mat L_k^T \mat L_k,
\eea
where $\mat L_k$ is a lower triangular matrix.
Any global unitary transformation
\bea
\hspace{-1cm}\mat U(\theta,\phi,\psi) & = & \big(\begin{smallmatrix}e^{\inumb\phi}&0\\0&e^{-\inumb\phi}\end{smallmatrix}\big)\big(\begin{smallmatrix}\cos\theta&-\sin\theta\\\sin\theta&\phantom{-}\cos\theta\end{smallmatrix}\big)\big(\begin{smallmatrix}e^{\inumb\psi}&0\\0&e^{-\inumb\psi}\end{smallmatrix}\big) \label{eq:unit}
\eea
of the diabatic potential [$\mat V=(\begin{smallmatrix}V_D&V_c\\V_c&V_A\end{smallmatrix})$, \eq{eq:QVC}] 
leads to a different diabatic model $\tilde{\mat V}=\mat U^\dagger\mat V\mat U$, 
which gives back the same adiabatic energies and the same quantum dynamics.

Since the diabatic state potentials must be bounded in any representation, 
the Hessians of the diabatic states must be positive definite for any value of 
$\phi$, $\theta$, and $\psi$. 
This positivity imposes some constraints on possible values of $K_c$ elements. 
All possible Hessians generated by the transformation can be written as
\bea\nonumber
\mat K_\theta & = & ( \mat K_A + \mat K_D ) + \cos(2\theta)( \mat K_A - \mat K_D ) \\
&& \hspace{1.95cm} + 2\cos(\phi)\sin(2\theta)\mat K_c,\label{eq:general-hess}
\eea
where $\theta$ and $\phi$ are parameters of the unitary transformation of the diabatic states. 
A parametrization of $\mat K_c$ that ensures the positivity of the Hessian 
given in \eq{eq:general-hess} is still an open problem.~\cite{Horn:1962/pjm/225,Fulton:2000/bams/209} 
We can fix $\phi$ to $0$ because it does not affect $K_c$ parametrization. 
Indeed, if for a given $\theta$ $\mat K_\theta$ is positive for $\cos(\phi)=\pm1$, 
then it is also positive for $-1<\cos(\phi)<1$ because the highest absolute 
eigenvalue of the matrix $\cos(\phi)\mat K_c$ can only be smaller.
The Cholesky decompositions of $\mat K_D$ and $\mat K_A$ already impose 
positivity for the cases $\theta=0,\pi/2$.
Positivity for $\theta=\pm\pi/4$ ($\mat K_\theta=\mat K_A + \mat K_D \pm 2\mat K_c$) 
can be ensured by using the following parametrization~\cite{Auyeung:1969/pams/545}
\bea\label{eq:Kc-param1}
\mat K_c & = & \frac{1}{2}\mat L_s^T e^{-\mat A} \diag(atan(\mat r)) e^{\mat A} \mat L_s,
\eea
where $\mat L_s$ is the Cholesky decomposition of $\mat K_A + \mat K_D$, $\mat A$ is 
a real anti-symmetric matrix, and $\mat r$ is a real-valued vector.

The parametrization given in \eq{eq:Kc-param1} is used in the \glspl{pm} parameter optimization. 
However, for optimizations of the full-dimensional models this parametrization becomes more 
difficult to implement and cannot be combined with constraints unless we substitute the steepest 
descent algorithm for another more computationally expensive one. 
To avoid this extra computational cost, we change the parametrization of 
$\mat K_c$ for this latter case to ensure positivity when $\theta=\pi/4$ 
($\mat K_\theta=\mat K_A + \mat K_D + 2\mat K_c$):
\bea\label{eq:Kc-param2}
\mat K_c & = & \frac{1}{2}\left(\mat L_c^T\mat L_c - \mat K_A - \mat K_D \right),
\eea
where the matrix $\mat L_c$ represents a set of parameters to be optimized for changing $\mat K_c$.

There are values of $\theta$ for which the positivity of $\mat K_\theta$ cannot be ensured by 
the parametrization. In these cases, we build a grid in the range $]-\pi/2:\pi/2]$ on which we check 
the positivity of $\mat K_\theta$.

\section{Optimization of the full-dimensional model's parameters}
\label{app:steep}

Here we describe our method of obtaining the parameters for the full-dimensional \gls{QVC} models.
The idea is based on the following expression for adiabatic energies of 
a two-electronic state system
\bea
E_\pm & = & (\Sigma \pm \Omega)/2 \label{eq:adiabenergies}
\eea
where $E_\pm$ are the adiabatic energies for lower, $S_1$, and upper, $S_2$, electronic states, $\Sigma$ is the sum of adiabatic energy, $\Omega$ is the absolute adiabatic energy difference.
$\Omega$ is non-linear with respect to the diabatic elements given by
\bea
\Omega^2 & = & \Delta^2+4V_c^2 \label{eq:Omega}
\eea
where $\Delta=V_A-V_D$ is the diabatic energy difference and $V_c$ is the diabatic coupling between diabatic electronic states [see \eq{eq:QVC}].
Equation~\eqref{eq:adiabenergies} suggests to fit the sum and the energy difference separately 
such that the sum part (quantities $\mat\nabla\Sigma$, $\mat\nabla\mat\nabla^T\Sigma$ at $M_0$, $M_D$ and $M_A$) can be fitted linearly to obtain full dimensional $\mat v_D+\mat v_A$, and $\mat K_D+\mat K_A$.
Because \eq{eq:Omega} is non-linear, one can employ a non-linear fitting procedure to obtain diabatic parameters.
However, it is possible to avoid this difficulty after inspecting 
differential forms of \eq{eq:Omega} with respect to nuclear coordinates
\bea\label{eq:gradient}
\Omega\mat\nabla\Omega & = & \Delta\mat\nabla\Delta + 4V_c\mat\nabla V_c,\\\label{eq:hessian}
(\mat\nabla\Omega)(\mat\nabla\Omega)^T + \Omega\mat\nabla\mat\nabla^T\Omega & = & (\mat\nabla\Delta)(\mat\nabla\Delta)^T\\\nonumber
&& \hspace{-3cm} + \Delta\mat\nabla\mat\nabla^T\Delta + 4(\mat\nabla V_c)(\mat\nabla V_c)^T + 4V_c\mat\nabla\mat\nabla^TV_c.
\eea
Because $M_0$, $M_D$ and $M_A$ are already described by the \gls{5D} model,
$\Delta$ and $V_c$ and especially the quantities $\{\mat v_k^T\mat x_j,\mat x_i^T\mat K_k\mat x_j; k=D, A, c\}$, where $\mat x_i$ and $\mat x_j$ are already known at the coordinate vectors for nuclear configurations $i, j=M_0, M_D, M_A$. 
Knowing $\Delta$ and $V_c$ at the minima makes possible to turn \eq{eq:gradient} and \eq{eq:hessian} 
at the stationary geometries in a set of linear equations with respect to the parameters $\{\mat v_k\}$ and $\{\mat K_k\}$. The system of linear equations is generated in two steps:
First, we project \eq{eq:hessian} on the stationary geometries and combine the resulting equations with \eq{eq:gradient}. 
The resulting system of equations can be easily solved as a least square problem to obtain the following linear terms: $\{\mat v_k,\mat K_k\mat x_j; k=D, A, c; j=M_0, M_D, M_A\}$.
Second, we obtain the quadratic terms of the model by substituting linear terms in \eq{eq:hessian}, which becomes linear with respect to matrices $\{\mat K_k\}$.

The special parametrization of the matrices $\{\mat K_k\}$ given by Eqs.~\eqref{eq:K-param} and~\eqref{eq:Kc-param2} 
makes a set of equations for $\{\mat K_k\}$ non-linear. 
Therefore, we minimize the corresponding least square problem with the
 steepest descent algorithm, which requires algebraic manipulations 
 with matrices of only the $\mat K_k$ size, and thus, is feasible and fast.
The minimization is constrained to ensure that we recover the accurate \gls{5D} models after projecting the full-dimensional models on the \glspl{pm}.

\bibliography{CPD-ET}

\begin{thebibliography}{47}
\bibitem [{mis()}]{misc:coupling}%
  \BibitemOpen
  \href@noop {} {\enquote {\bibinfo {title} {The relative coupling between the
  two directions $j$ and $k$ is given by $2g_{K,jk}/(g_{K,jj}-g_{K,kk})$, where
  $g_{K,jk}$ is the element of the adiabatic hessian matrix in the
  $k^\text{th}$ electronic state.}}\ }\BibitemShut {NoStop}%
\end{thebibliography}
\end{document}